\documentstyle[12pt,a4,epsfig]{article}
\advance\topmargin by -1.2cm
\newcommand{\be}{\begin{equation}}
\newcommand{\ee}{\end{equation}}
\newcommand{\bea}{\begin{eqnarray}}
\newcommand{\eea}{\end{eqnarray}}

\newcommand{\bm}[1]{\mbox{\boldmath $#1$}}

\newcommand{\xbj}{x_{\scriptscriptstyle B}}

\def\st{{\scriptscriptstyle T}}
\def\slash{\rlap{/}}

\begin{document}

\centerline{\Large \bf THE FULL SPIN STRUCTURE}
\vskip 0.2 cm
\centerline{\Large \bf OF QUARKS IN THE NUCLEON\footnote{
Talk presented by P.J. Mulders at the
workshop on Nucleon Structure in the High x-Bjorken Region
(HiX2000), Temple University, Philadelphia, PA, USA; March 30 - April 1,
2000}
}
\vskip 0.5 cm
\centerline{A. Bacchetta, M. Boglione, A. Henneman and
P.J. Mulders}
\vskip 0.3 cm
\centerline{\small\it
Department of Theoretical Physics, Faculty of Science,
Vrije Universiteit}
\vskip 0.1 cm
\centerline{\small\it
De Boelelaan 1081, NL-1081 HV Amsterdam, the Netherlands}
\vskip 1 cm

\begin{abstract}
We discuss bounds on the 
distribution and fragmentation functions that
appear at leading order in deep inelastic 1-particle inclusive
leptoproduction or in Drell-Yan processes. These bounds simply follow
from positivity of the quark-hadron scattering matrix elements and are an 
important guide in estimating the magnitude of the azimuthal
and spin asymmetries in these processes. We focus on an example
relevant for deep inelastic scattering at relatively low energies.
\end{abstract}

\section{The spin structure in inclusive DIS}

\begin{figure}[t]
\begin{center}
\epsfig{file=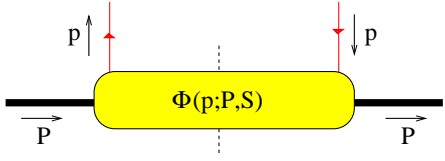,width = 6.0 cm}
\end{center}
\caption{\label{fig1}
Matrix element for distribution functions.}
\end{figure}

In deep-inelastic scattering (DIS) the transition from hadrons to 
quarks and gluons is described in terms of distribution and
fragmentation functions.
In general, the distribution functions for a quark can be
obtained from the lightcone correlation 
functions~\cite{Soper77,Jaffe83,Manohar90,JJ92}.  
\be
\Phi_{ij}(x) = \left. \int \frac{d\xi^-}{2\pi}\ e^{ip\cdot \xi}
\,\langle P,S\vert \overline \psi_j(0) \psi_i(\xi)
\vert P,S\rangle \right|_{\xi^+ = \xi_\st = 0},
\ee
depending on the lightcone fraction $x = p^+/P^+$.
The hadron momentum $P$ is chosen so that 
it has no transverse component, $P_\st = 0$.
At leading order, the relevant part of the correlator is $\Phi\gamma^+$ 
%Inserting a complete set of intermediate states and generalizing to
%off-diagonal spin, one obtains
\bea
(\Phi\gamma^+)_{ij}
& = & \left. \int \frac{d\xi^-}{2\pi\sqrt{2}}\ e^{ip\cdot \xi}
\,\langle P,s^\prime\vert \psi^\dagger_{+j}(0) \psi_{+i}(\xi)
\vert P,s\rangle \right|_{\xi^+ = \xi_\st = 0}
%\nonumber \\
%& = & \frac{1}{\sqrt{2}}\sum_n
%\langle P_n\vert \psi_{+j}(0)\vert P,s^\prime\rangle^\ast
%\langle P_n\vert \psi_{+i}(0)\vert P,s\rangle
%\,\delta\left(P_n^+ - (1-x)P^+\right) ,
\label{dens}
\eea
where $\psi_+ \equiv P_+\psi = \frac{1}{2}\gamma^-\gamma^+\psi$ is the
good component of the quark field~\cite{KS70}. \\
The correlator contains all the soft parts appearing in the scattering
processes and, as shown in Fig.~\ref{fig1}, is related to the 
forward amplitude for antiquark-hadron scattering. By considering the
quantity $M=(\Phi\gamma^+)^T$, 
one finds that for any antiquark-hadron state $|a\rangle$ the expectation 
value $\langle a|M|a\rangle$  must be larger than or equal to zero.

Thus our strategy is the following: express the forward scattering matrix M 
as a matrix in the [parton chirality space $\otimes$ hadron spin space] 
and obtain our bounds by requiring it to be positive semi-definite. 

At leading twist and when no partonic intrinsic transverse 
momentum is taken into account, $\Phi(x) \gamma^+$ is simply given by the 
contribution of three distribution functions~\cite{remark}
%,  the well known $f_1(x)$ and 
%$g_1(x)$, and the chiral odd and still experimentally unmeasured $h_1(x)$
\be
\Phi(x) \gamma^+ = \Bigl\{
f_1(x) + \lambda\,g_1(x)\,\gamma_5  + h_1(x)\,\gamma_5\,\slash S_\st
\Bigr\}P_+ \,.
\label{phiint}
\ee
The first step consists in writing this quantity  
as a matrix in the {\it parton} chirality space; this is easily done by 
using the explicit expression of 
$P_+ = \frac{1}{2} \gamma ^- \gamma ^+$ and $\gamma _5$ 
as $4\times 4$ Dirac matrices. 
In chiral representation we find
\be
M_{ij} =
\left\lgroup \begin{array}{cccc}
f_1(x) + \lambda\,g_1(x) & 0 & 0 & (S_\st^1+i\,S_\st^2)\,h_1(x) \\
0 & 0 & 0 & 0 \\
0 & 0 & 0 & 0 \\
(S_\st^1-i\,S_\st^2)\,h_1(x) & 0 & 0 & f_1(x) - \lambda\,g_1(x)
\end{array}\right\rgroup
\ee
As it is clear from the above expression, at leading twist  
 there are only two relevant basis states, corresponding to 
(good components of) the right and the left handed partons. Thus, instead of 
using the full four dimensional Dirac space, we can effectively use only a 
two-dimensional chirality space. 

Step two will be to write the above matrix explicitly in the {\it hadron} 
spin space. 
In order to study the correlation function in a spin $1/2$ target we introduce
 a spin vector $S$ that parameterizes the spin density matrix 
\be
\rho (P,S)=\frac{1}{2}( 1 + \bm{S\cdot \sigma}).
\ee 
The spin vector satisfies
$P\cdot S$ = 0 and $S^2 = -1$ (spacelike) for a pure state, $-1 <
S^2 \le 0$ for a mixed state. Using $\lambda \equiv MS^+/P^+$ and 
the transverse spin vector $S_\st$, the condition becomes 
$\lambda^2 + \bm S_\st^2 \le 1$, as can be seen from the rest-frame
expression $S$ = $(0,\bm S_\st, \lambda)$. The precise equivalence
of a $2\times 2$ matrix
$\tilde M_{ss^\prime}$ in the target spin space
and the $S$-dependent function $M(S)$ is
$M(S) = \mbox{Tr}\left[ \rho(S)\,\tilde M\right]$.
Explicitly, the $S$-dependent function
\be
M(S) = M_O + \lambda\,M_L + S_\st^1\,M_\st^1
+ S_\st^2\,M_\st^2,
\label{M(S)}
\ee
corresponds to a matrix, which
in the target rest-frame with as basis the spin 1/2 
states with $\lambda = +1$ and $\lambda = -1$ becomes
\be
\tilde M_{ss^\prime} =
\left\lgroup \begin{array}{cc}
M_O + M_L & M_\st^1 - i\,M_\st^2 \\
& \\
M_\st^1 + i\,M_\st^2 & M_O - M_L \\
\end{array}\right\rgroup
\label{spinexplicit}
\ee
Thus, each element of matrix (4) will transform in a $2\times 2$ matrix, 
according to Eqs.(\ref{M(S)}) and (\ref{spinexplicit}).

At leading twist and in absence of intrinsic transverse momentum, in the combined [parton chirality $\otimes$ hadron 
spin  space], the final 
result is
\be
\tilde M _{is,js\prime}=
\left\lgroup \begin{array}{cccc}
f_1 + g_1 & 0 & 0 & 2\,h_1 \\
& \\
0 & f_1 - g_1 & 0 & 0 \\
& \\
0 & 0 & f_1 - g_1 & 0 \\
& \\
2\,h_1 & 0 & 0 & f_1 + g_1
\end{array}\right\rgroup .
\ee
From the
positivity of the diagonal elements one recovers the trivial bounds
$f_1(x) \ge 0$ and $\vert g_1(x) \vert \le f_1(x)$, but
requiring the eigenvalues of the matrix to be positive gives
the stricter Soffer bound~\cite{Soffer73},
\be
\vert h_1(x)\vert \le \frac{1}{2}\left( f_1(x) + g_1(x)\right) .
\ee

\section{The full spin structure in SIDIS}

We now turn to the more general case in which non-collinear configurations 
are taken into account. Transverse momenta of the partons inside the proton 
play an important role in hard processes 
with more than one hadron~\cite{RS79}, like semi-inclusive deep inelastic 
scattering (SIDIS), $e^- H \rightarrow e^- h X$~\cite{MT96}, 
or Drell-Yan scattering, $H_1 H_2 \rightarrow \mu^+\mu^- X$~\cite{TM95}.

Analogous bounds can be obtained for transverse momentum 
dependent distribution and fragmentation functions.
The soft parts involving the distribution functions are
contained in the lightfront correlation function
\be
\Phi_{ij}(x,\bm p_T) =
\left. \int \frac{d\xi^-d^2\bm \xi_T}{(2\pi)^3}\ e^{ip\cdot \xi}
\,\langle P,S\vert \overline \psi_j(0) \psi_i(\xi)
\vert P,S\rangle \right|_{\xi^+ = 0},
\ee
depending on $x=p^+/P^+$ and the quark transverse 
momentum $\bm p_\st$ in a target with $P_\st = 0$. 

Separating the terms corresponding to unpolarized ($O$), 
longitudinally polarized ($L$) and transversely polarized targets
($T$), the most general parameterizations with $p_\st$-dependence, 
relevant at leading order, are
\bea
\Phi_O(x,\bm p_\st)\,\gamma^+ & = &
\Biggl\{
f_1(x,\bm p^2_\st)
+ i\,h_1^\perp(x,\bm p^2_\st)\,\frac{\slash p_\st}{M}
\Biggr\} P_+\,,\nonumber
\\
\Phi_L(x,\bm p_\st)\,\gamma^+ & = &
\Biggl\{
\lambda\,g_{1L}(x,\bm p^2_\st)\,\gamma_5
+ \lambda\,h_{1L}^\perp(x,\bm p^2_\st)
\gamma_5\,\frac{\slash p_\st}{M}
\Biggr\} P_+\,,\nonumber
\\
\Phi_T(x,\bm p_\st)\,\gamma^+  & = &
\Biggl\{
f_{1T}^\perp(x,\bm p^2_\st)\,\frac{\epsilon_{\st\,\rho \sigma}
p_\st^\rho S_\st^\sigma}{M}
+ g_{1T}(x,\bm p^2_\st)\,\frac{\bm p_\st\cdot\bm S_\st}{M}
\,\gamma_5
\nonumber \\ & &\mbox{}
+ h_{1T}(x,\bm p^2_\st)\,\gamma_5\,\slash S_\st
+ h_{1T}^\perp(x,\bm p^2_\st)\,\frac{\bm p_\st\cdot\bm S_\st}{M}
\,\frac{\gamma_5\,\slash p_\st}{M}
\Biggr\} P_+\,,
\label{Phi-full}
\eea
where indeed $\Phi(x,\bm p_\st)=\Phi_O(x,\bm p_\st)+\Phi_L(x,\bm p_\st)+
\Phi_T(x,\bm p_\st)$.
As before, $f_{\ldots}$, $g_{\ldots}$ and $h_{\ldots}$
indicate unpolarized, chirality and transverse spin distributions.
The subscripts $L$ and $T$
indicate the target polarization, and the superscript $\perp$ signals 
explicit presence of transverse momentum of partons.
Using the notation $f^{(1)}(x,\bm p^2_\st) \equiv 
(\vert \bm p_\st\vert^2/2M^2)\,f(x,\bm p^2_\st)$,
one sees that $f_1(x,\bm p_\st^2)$, $g_1(x,\bm p_\st^2) = g_{1L}(x,\bm p_\st^2)$ and $h_1(x,\bm p_\st^2) = h_{1T}(x,\bm p_\st^2) + h_{1T}^{\perp (1)}(x,\bm p_\st^2)$ are the functions surviving $p_\st$-integration.

To put bounds on the transverse momentum dependent functions, we
again make the matrix structure explicit, following the same procedure we 
used in the previous simpler case in which no $p_T$ was taken into account. 
We find for $M = (\Phi(x,\bm p_\st)\,\gamma^+)^T$
the full spin matrix (for simplicity we do not explicitly indicate the 
$x$ and $\bm p_\st^2$ dependence of the distribution functions)
\[\tilde M=
\left\lgroup \begin{array}{cccc}
f_1 + g_{1L} &
\frac{\vert p_\st\vert}{M}\,e^{i\phi}\,g_{1T}&
\frac{\vert p_\st\vert}{M}\,e^{-i\phi}\,h_{1L}^\perp&
2\,(h_{1T} + h_{1T}^{\perp (1)}) \\
& & & \\
\frac{\vert p_\st\vert}{M}\,e^{-i\phi}\,g_{1T}^\ast&
f_1 - g_{1L} &
\frac{\vert p_\st\vert^2}{M^2}e^{-2i\phi}\,h_{1T}^\perp &
-\frac{\vert
p_\st\vert}{M}\,e^{-i\phi}\,h_{1L}^{\perp\ast}\\ & & &
\\ \frac{\vert p_\st\vert}{M}\,e^{i\phi}\,h_{1L}^{\perp\ast}&
\frac{\vert p_\st\vert^2}{M^2}e^{2i\phi}\,h_{1T}^\perp &
f_1 - g_{1L} &
-\frac{\vert p_\st\vert}{M}\,e^{i\phi}\,g_{1T}^\ast \\
& & & \\
 2\,(h_{1T} + h_{1T}^{\perp (1)})&
-\frac{\vert p_\st\vert}{M}\,e^{i\phi}\,h_{1L}^\perp&
-\frac{\vert p_\st\vert}{M}\,e^{-i\phi}\,g_{1T} &
f_1 + g_{1L}
\end{array}\right\rgroup 
\]
where $\phi$ is the azimuthal angle of the transverse momentum vector.
Here, we have left out the T-odd functions. But time-reversal invariance was 
not imposed in the parameterization of $(\Phi(x,\bm p_\st)\,\gamma^+)$ in 
Eqs.~(\ref{Phi-full}), allowing for non-vanishing T-odd functions 
$f_{1T}^\perp(x,\bm p^2_\st)$ and $h_1^\perp(x,\bm p^2_\st)$. 
They can be easily incorporated
as the imaginary parts of the functions $g_{1T}(x,\bm p^2_\st)$ and 
$h_{1L}^\perp(x,\bm p^2_\st)$,
to be precise $g_{1T} \rightarrow g_{1T}+i\,f_{1T}^\perp$ and 
$h_{1L}^\perp \rightarrow h_{1L}^\perp+i\,h_1^\perp$.
Possible sources of T-odd effects in the initial state have been 
discussed in Refs~\cite{Sivers90}.

This matrix is particularly relevant, as it illustrates the 
full quark spin structure accessible in a polarized nucleon~\cite{BoM99}, 
which is equivalent to the full spin structure of the forward 
antiquark-nucleon scattering amplitude.
Bounds to insure positivity of any matrix element can be
obtained by looking at the 1-dimensional and 2-dimensional subspaces and at 
the eigenvalues of the full matrix.
The 1-dimensional subspaces give the trivial bounds 
\be
f_1(x,\bm p_\st^2) \ge 0 \, , 
\ee
\be
\vert g_{1L}(x,\bm p_\st^2)\vert \le f_1(x,\bm p_\st^2) \,.
\ee
From the 2-dimensional subspaces we get
\bea
&& \vert h_1 \vert \le
\frac{1}{2}\left( f_1 + g_{1L}\right)
\le f_1,
\label{Soffer}\\
&&
\vert h_{1T}^{\perp(1)}\vert \le
\frac{1}{2}\left( f_1 - g_{1L}\right)
\le f_1,
\\
&& \vert g_{1T}^{(1)}\vert^2
\le \frac{\bm p_\st^2}{4M^2}
\left( f_1 + g_{1L}\right)
\left( f_1 - g_{1L}\right)
\le \frac{\bm p_\st^2}{4M^2}\,f_1^2,
\\
&& \vert h_{1L}^{\perp (1)}\vert^2
\le \frac{\bm p_\st^2}{4M^2}
\left( f_1 + g_{1L}\right)
\left( f_1 - g_{1L}\right)
\le \frac{\bm p_\st^2}{4M^2}\,f_1^2,
\eea
where, once again, we did not explicitly indicate the $x$ and $\bm p_\st^2$ 
dependence to avoid too heavy a notation.
Besides the Soffer bound, Eq.~(\ref{Soffer}), new bounds
for the distribution functions are found. In particular, one sees that 
functions like
$g_{1T}^{(1)}(x,\bm p_\st^2)$ and $h_{1L}^{\perp (1)}(x,\bm p_\st^2)$ 
appearing in azimuthal asymmetries
in leptoproduction are proportional to $\vert \bm p_\st\vert$ for small
$p_\st$. 

Before sharpening these bounds via the eigenvalues, it is convenient to
introduce two positive definite functions $A(x,\bm p_\st^2)$ 
and $B(x,\bm p_\st^2)$ such that
$f_1 = A + B$ and $g_1 = A - B$ and define
\bea
&&
h_1(x,\bm p_\st^2) = \alpha \,A ,
\\ &&
h_{1T}^{\perp (1)}(x,\bm p_\st^2) = \beta\,B ,
\\ &&
g_{1T}^{(1)}(x,\bm p_\st^2) =
\gamma \,\frac{\vert p_\st\vert}{M}
\,\sqrt{A\,B} ,
\\ &&
h_{1L}^{\perp (1)}(x,\bm p_\st^2) =
\delta \,\frac{\vert p_\st\vert}{M}
\,\sqrt{A\,B} ,
\eea
where the functions $\alpha(x,\bm p_\st^2)$, $\beta(x,\bm p_\st^2)$, 
$\gamma(x,\bm p_\st^2)$ and $\delta(x,\bm p_\st^2)$ 
have absolute values in the interval $[-1,1]$. Note that
$\alpha$ and $\beta$ are real-valued but
$\gamma$ and $\delta$ are complex-valued, the imaginary part
determining the strength of the T-odd functions. 

\begin{figure}[t]
\begin{center}
\epsfig{file=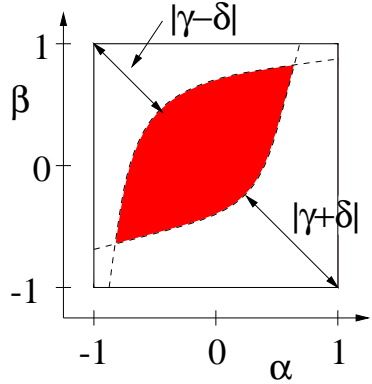,width = 5.0 cm}
\end{center}
\caption{\label{fig2}
Allowed region (shaded) for $\alpha$ and $\beta$ depending on
$\gamma$ and $\delta$.}
\end{figure}

Next we sharpen these bounds using the eigenvalues of the
matrix, which are given by
\bea
e_{1,2} = (1-\alpha)A + (1+\beta)B
\pm \sqrt{4AB\vert\gamma+\delta\vert^2+((1-\alpha)A-(1+\beta)B)^2},
\\
e_{3,4} = (1+\alpha)A + (1-\beta)B
\pm \sqrt{4AB\vert\gamma-\delta\vert^2+((1+\alpha)A-(1-\beta)B)^2}.
\eea
Requiring them to be positive can be converted into the conditions
\bea
&&
A+B\ge 0.
\\ &&
\vert \alpha\,A-\beta\,B \vert \le A+B,
\quad \mbox{i.e.} \ \vert h_{1T}(x,\bm p_\st^2)\vert \le f_1(x,\bm p_\st^2)
\\ &&
\vert \gamma + \delta \vert^2 \le (1-\alpha)(1+\beta) ,
\\ &&
\vert \gamma - \delta \vert^2 \le (1+\alpha)(1-\beta) .
\eea
It is interesting for the phenomenology of deep inelastic processes that 
a bound for the transverse spin distribution $h_1$ is provided 
not only by the inclusively measured functions $f_1$ and 
$g_1$, but also
by the functions $g_{1T}(x,\bm p_\st^2)$ and 
$h_{1L}^{\perp}(x,\bm p_\st^2)$, 
responsible for specific azimuthal asymmetries~\cite{MT96,BM98}.
This is illustrated in Fig.~\ref{fig2}. 

\begin{figure}[t]
\begin{center}
\epsfig{file=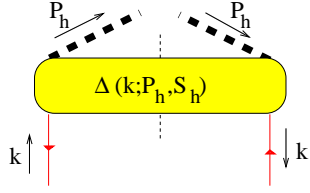,width = 4.0 cm}
\end{center}
\caption{\label{fig3}
Matrix element for fragmentation functions.}
\end{figure}

A perfectly analogous calculation con be performed for fragmentation 
functions, which describe the hadronization process of a parton into the 
final detected hadron. In this case the transverse momentum dependent 
correlator is~\cite{CS82}
\be
\Delta_{ij}(z,\bm k_\st) =
\left. \sum_X \int \frac{d\xi^-d^2\bm \xi_\st}{(2\pi)^3} \,
e^{ik\cdot \xi} \langle 0 \vert \psi_i (\xi) \vert P_h,X\rangle
\langle P_h,X\vert\overline \psi_j(0) \vert 0 \rangle
\right|_{\xi^+ = 0},
\ee
(see Fig.~\ref{fig3}) 
depending on $z = P_h^+/k^+$ and the quark transverse momentum $k_\st$ 
leading to a hadron with $P_{h\st} = 0$. A simple boost shows 
that this is equivalent to a quark producing a hadron with transverse 
momentum $P_{h\perp} = -z\,k_\st$ with respect to the quark.

Like $\Phi$, $\Delta$ is parameterized in terms of 
unpolarized, chirality and transverse-spin fragmentation 
functions~\cite{MT96},
denoted by capital letters $D_{\ldots}$, $G_{\ldots}$, and $H_{\ldots}$, 
respectively.
For the fragmentation process, time-reversal invariance cannot be 
imposed~\cite{RKR71,HHK83,JJ93}, and the T-odd fragmentation 
functions $D_{1T}^{\perp}$~\cite{MT96} and 
$H_1^\perp$~\cite{Collins93} 
play a crucial role in some azimuthal spin asimmetries as we shall discuss 
later on. 

All the bounds obtained for the distribution functions can be rephrased in 
terms of the corresponding fragmentation functions.
For instance, the relevant bounds for the Collins function 
$H_1^{\perp (1)}$, describing the fragmentation of a 
transversely polarized quark into a (spin zero) pion becomes
\be
H_1^{\perp (1)}(z_\pi,\bm P_{\pi\perp}^2 ) \le \frac{\vert 
\bm P_{\pi\perp}\vert}{2z_\pi M_\pi} D_1(z_\pi,\bm P_{\pi\perp}^2) \,,
\ee
while for the other T-odd function $D_{1T}^{\perp (1)}$, describing 
fragmentation of an unpolarized quark into a polarized hadron 
such as a $\Lambda$ one has
\be
D_{1T}^{\perp (1)}(z_\Lambda,\bm P_{\Lambda\perp}^2) \le \frac{\vert 
\bm P_{\Lambda\perp}\vert}{2z_\Lambda M_\Lambda} 
D_1(z,\bm P_{\Lambda\perp}^2) \,.
\ee
Similarly to what happened for the distribution functions, a bound for the 
transverse spin fragmentation $H_1^\perp$ is provided not 
only by the inclusive function $D_1$ but, when we sharpen 
the bounds by requiring positivity of the eigenvalues of the full matrix, 
the magnitude of $H_1^\perp$ also constrains the magnitude of 
$H_1$~\cite{J96}.

Recently SMC~\cite{SMC}, HERMES~\cite{HERMES} and LEP~\cite{LEP} 
have reported preliminary results for azimuthal asymmetries. More results
are likely to come in the next few years from HERMES, RHIC and 
COMPASS experiments. Although much theoretical work is needed, 
for instance on factorization, scheme ambiguities and 
the stability of the bounds under evolution~\cite{scheme}, 
these future experiments may provide us with the knowledge of the
full helicity structure of quarks in a nucleon. 
The elementary bounds derived in this paper can serve as 
important guidance to estimate the magnitudes of 
asymmetries expected in the various processes.

\section{An example, relevant for JLAB@12 GeV}

The asymmetries for which evidence recently has been found are 
mostly  single spin asymmetries involving T-odd fragmentation  
functions, such
as the Collins function $H_1^\perp$. We would like to discuss here a
measurement that can be combined with the measurement of the
inclusive structure function $g_2(x)$ in deep inelastic scattering off a
transversely polarized target at large $x$. 
%and for large $x$ measurements are 
%important.

\begin{figure}[t]
\begin{center}
\begin{minipage}{8.5cm}
\epsfig{file=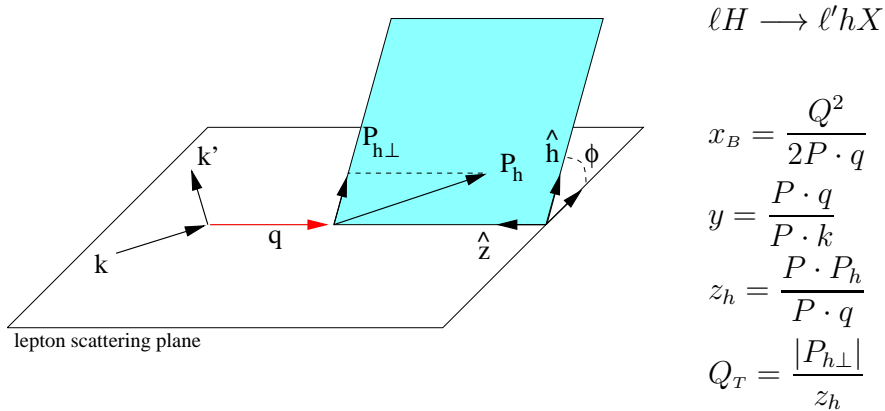,width=8.5cm}
\end{minipage}
\begin{minipage}{2.5cm}
\begin{eqnarray*}
&&\ell H \longrightarrow \ell^\prime h X
\end{eqnarray*}
\begin{eqnarray*}
&&\xbj = \frac{Q^2}{2P\cdot q} \\
&&y = \frac{P\cdot q}{P\cdot k}\\
&&z_h = \frac{P\cdot P_h}{P\cdot q}\\
&&Q_\st = \frac{\vert P_{h\perp}\vert}{z_h}
\end{eqnarray*}
\end{minipage}
\end{center}
\caption{\label{fig4}
Kinematics for 1-particle inclusive leptoproduction.}
\end{figure}

The relevant kinematic variables are illustrated 
in Fig.~\ref{fig4}, where also the scaling
variables are introduced.
Most often one considers the cross section integrated over all
transverse momenta. But we emphasize that in principle the
cross section can depend on a transverse vector, for which we
use $\bm q_\st$ = $-P_{h\perp}/z_h$. This vector either represents
the transverse momentum of the photon momentum $q$ (with respect
to the two hadrons, target and produced hadron) or the transverse
momentum of the produced hadron (with respect to the target and
photon momenta). Introducing the weighted cross sections 
\bea
&&
\left< W \right>_{P_eP_HP_h}
\equiv
\int d\phi^\ell\,d^2\bm q_{T}
\,W(Q_\st,\phi_h^\ell,\phi_S^\ell,\phi_{S_h}^\ell)
\,\frac{d\sigma_{{P_eP_HP_h}}}{d\xbj\,dy\,dz_h\,d\phi^\ell\,d^2\bm q_{T}},
\eea
where $W$ is some weight depending on azimuthal angles and transverse
momentum and the subscripts $P_e$, $P_H$ and $P_h$ are the 
polarizations of lepton, target and produced hadron respectively, 
we can construct several asymmetries.

To illustrate the weights, let's consider an easy example: the standard 
$\bm q_\st$-integrated 1-particle inclusive unpolarized cross section,
\be
\frac{d\sigma_{OO}}{d\xbj\,dy\,dz_h}
= \frac{2\pi \alpha^2\,s}{Q^4}\,\sum_{a,\bar a} e_a^2
\left\lgroup 1 + (1-y)^2\right\rgroup \xbj {f^a_1}(\xbj)\,{ D^a_1}(z_h),
\ee
in this language becomes 
\be
\left< 1 \right>_{OO}
= \frac{2\pi \alpha^2\,s}{Q^4}\,\sum_{a,\bar a} e_a^2
\left\lgroup 1 + (1-y)^2\right\rgroup \xbj {f^a_1}(\xbj)\,{ D^a_1}(z_h).
\ee
One of the leading asymmetries involving the function 
$g_{1T}(x,\bm p_\st^2)$ discussed in the previous section 
is an asymmetry for longitudinally polarized leptons off a transversely
polarized nucleon~\cite{KT}
\be
\left< \frac{Q_\st}
{M} \,\cos(\phi_h^\ell-\phi_S^\ell)\right>_{LT}
= \frac{2\pi \alpha^2\,s}{Q^4}\,{\lambda_e\,\vert \bm S_\st \vert}
\,y(2-y)\sum_{a,\bar a} e_a^2
\,\xbj\,{g_{1T}^{(1)a}}(\xbj) {D^a_1}(z_h),
\label{proposed}
\ee
Since the fragmentation function involved is the standard leading
one for unpolarized quarks into unpolarized or spin 0 hadrons, one
can consider it for pion production, for which the fragmentation
functions are reasonably well-known~\cite{TMR}.

\begin{figure}[t]
\begin{center}
\epsfig{file=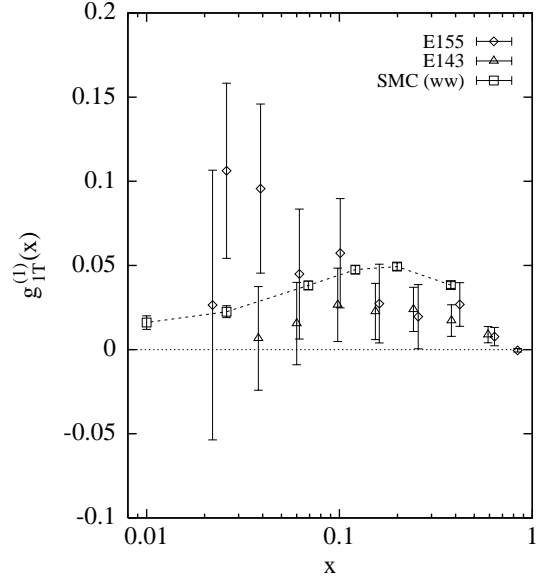,width=7.5cm}
\end{center}
\caption{\label{fig5}
Estimate of $g_{1T}(x)$ obtained from $g_2$ data or from the
Wandzura-Wilczek approximation using the SMC $g_1$-data
(see ref.~\cite{BM00} for details).}
\end{figure}

As we mentioned before, it is interesting to do this measurements
together with the $g_2$-measurement which, expressed as an 
(inclusive) asymmetry is given by
\be
\left< \cos\phi_S^\ell\right>_{LT} = 
-\lambda_e\,\vert\bm S_\st\vert
\,y\,\sqrt{1-y} \ \sum_{a,\bar a} e_a^2
\frac{M\xbj^2}{Q}\,g_T^a(\xbj) 
\ee
where $g_T^a(x) = g_1^a(x) + g_2^a(x)$. 
The comparison of the inclusive measurement of $g_2$ and the 
semi-inclusive measurement of $g_{1T}^{(1)}$ would enable one 
to test the relation~\cite{BKL,MT96,KM},
\be
g_2^a(x) = \frac{d}{dx}\,g_{1T}^{(1)a}.
\ee
relating a twist three function to a transverse momentum dependent
function. At present this relation can be used to estimate
the function  $g_{1T}(x)$ from existing $g_2$ measurements,
of course in the same flavor averaged way as an
inclusive measurement allows. The result is shown in
Fig.~\ref{fig5}, taken from ref.~\cite{BM00}. A measurement of
the asymmetry in
Eq.~\ref{proposed} allows an independent measurement of $g_{1T}$
and a test of the above relation.

\vspace{0.5cm}

\section*{Acknowledgments}

We would like to thank Elliot Leader for useful discussions.
This work is part of the research program of the
Foundation for Fundamental Research on Matter (FOM)
and the Dutch Organization for Scientific Research (NWO).
It is also part of the TMR program ERB FMRX-CT96-0008.


\begin{thebibliography}{30}
\bibitem{remark}
Other common notations for a quark flavor $q$ are $f^q_1(x) = q(x)$, 
$g^q_1(x) = \Delta q(x)$ and $h^q_1(x) = \Delta_\st q(x)$
\end{thebibliography}
\end{document}